# Strong coupling of monolayer $WS_2$ excitons and surface plasmon polaritons in a planar Ag/$WS_2$ hybrid structure

Nicolas Zorn Morales[1], Daniel Steffen Rühl[1], Sergey Sadofev[2], Giovanni Ligorio[1], Emil List-Kratochvil[3], Günter Kewes[4] and Sylke Blumstengel[1]*

**Affiliations**

[1]Department of Physics, Department of Chemistry & IRIS Adlershof, Humboldt-Universität zu Berlin, 12489 Berlin, Germany;

[2]Department of Physics, Humboldt-Universität zu Berlin, 12489 Berlin, German; present address: Institute of Crystal Growth, 12489 Berlin, Germany

[3]Department of Physics, Department of Chemistry & IRIS Adlershof, Humboldt-Universität zu Berlin, 12489 Berlin, Germany and Helmholtz-Zentrum Berlin für Materialien und Energie GmbH, Hahn-Meitner-Platz 1, 14109 Berlin, Germany

[4]Department of Physics & IRIS Adlershof, Humboldt-Universität zu Berlin, 12489 Berlin, Germany

*Corresponding author: sylke.blumstengel@physik.hu-berlin.de

**ABSTRACT**

Monolayer (1L) transition metal dichalcogenides (TMDC) are of strong interest in nanophotonics due to their narrow-band intense excitonic transitions persisting up to room temperature. When brought into resonance with surface plasmon polariton (SPP) excitations of a conductive medium opportunities for studying and engineering strong light-matter coupling arise. Here, we consider a most simple geometry, namely a planar stack composed of a thin silver film, an $Al_2O_3$ spacer and a monolayer of $WS_2$. We perform total internal reflection ellipsometry which combines spectroscopic ellipsometry with the Kretschmann-Raether-type surface plasmon resonance configuration. The combined amplitude and phase response of the reflected light at varied angle of incidence proves that despite the atomic thinness of 1L-$WS_2$, the strong coupling (SC) regime between A excitons and SPPs propagating in the thin Ag film is reached. The phasor representation of $\rho = r_\mathrm{p}/r_s$, where $r_\mathrm{p}$ and $r_\mathrm{s}$ are the Fresnel refection coefficients in p- and s-polarization, respectively corroborates SC as $\rho$ undergoes a topology change indicated by the occurrence of a double point at the cross over from the weak to the strong coupling regime. Our findings are validated by both analytical transfer matrix method calculations and numerical Maxwell simulations. The findings open up new perspectives for applications in plasmonic modulators and sensors benefitting from the tunability of the optical properties of 1L-TMDCs by electric fields, electrostatic doping, light and the chemical environment.

# 1 Introduction

Two-dimensional (2D) semiconducting transition metal dichalcogenides (TMDC) are currently of strong interest in nanophotonics due to their narrow-band intense excitonic transitions [1, 2]. When brought into resonance with electromagnetic fields in microcavities or plasmonic nanostructures unique opportunities for studying and engineering strong light-matter coupling arise [3-6]. Roughly speaking, the strong coupling (SC) regime is achieved whenever the energy exchange rate between two components is fast enough so that the energy can oscillate at least once back and forth. The resulting quasiparticles composed partly of light and partly of matter are known as polaritons. Polariton formation has been widely studied over the past decades using traditional III-V, II-VI as well as organic semiconductors as hosting materials which led to several remarkable breakthroughs, including polariton lasing [7], Bose–Einstein condensation [8] and superfluidity [9]. More recently, TMDC monolayers (1L) have emerged as promising candidates for the realization of strong coupling at room temperature which is difficult to achieve with traditional 3D inorganic semiconductors because of their small exciton binding energies. In contrast, due to the decreased screening and 2D confinement in 1L-TMDCs the electron-hole binding energies reach some 100 meV [10] and therefore, the excitons are stabile up to room temperature [1]. A further advantage is the tunability of the TMDC optical properties via external stimuli like electric fields, light, dielectric/chemical environment that provides the opportunity of active control over the light matter interaction which is prerequisite for several technological applications, such as sensing, ultra-fast optical switching, photon routing, energy/information transport in the nanometer length scale and possibly even information processing [11-14]. Triggered by these prospects, SC between excitons of 1L and multilayer TMDCs and cavity photons has been studied in all dielectric microcavities and in metal-based microcavities [5, 6, 15-18], however, mostly assisted by cryogenic cooling. Since the coupling strength is proportional to the inverse mode volume, plasmonic nanostructures or cavities featuring subwavelength mode confinement and strong field enhancement can provide a platform that performs under ambient conditions. SC was reported with localized surface plasmons in individual metallic nanostructures like Ag nanorods [19], Au nanorods [20, 21] and Ag nanoprisms [22-24] as well as in plasmonic nanoresonators [25, 26] at room temperature. Typically, a dip in the scattering spectrum is interpreted as a signature of the SC regime [22, 27, 28]. Controllable and reversible switching of the exciton-plasmon as well as trion-plasmon interaction was demonstrated via electro-static gating and temperature control [21, 29, 30].

Here, we follow an alternative route and study the coupling of 1L-TMDC excitons with surface plasmon polaritons (SPP) supported by a thin planar Ag film. Such a planar hybrid geometry is interesting for the realization active plasmonic components such as of all-optical or electro-optical plasmonic modulators [11, 14, 31]. So far, coupling of excitons in 1L-WSe$_2$ with SPP propagating in a planar Ag waveguide has been realized in combination with a plasmonic crystal cavity which was employed to enhance the coupling strength and to reach SC [14]. Until now, strong exciton-plasmon coupling in a planar geometry was only realized utilizing thick multilayer MoS$_2$ films [32]. Here, we demonstrate experimentally as well as theoretically that the SC regime under ambient conditions can even be achieved with a 1L of WS$_2$ in a planar geometry of ultimate simplicity, i.e. without the need of additional nanostructuring of the metal surface. Instead of the typically performed measurements of the intensity of the reflected or scattered light, we employ Total Internal Reflection Ellipsometry (TIRE) and show that this method provides a unique fingerprint to distinguish weak and SC regime. TIRE combines spectroscopic ellipsometry with the Kretschmann-Raether-type surface plasmon resonance (SPR) geometry [33]. The technique has been originally developed for biosensing applications since the simultaneous amplitude and phase measurement of the complex reflection ratio leads to a considerable sensitivity increase of the traditional SPR technique [34]. In the present context, the combined amplitude and phase response provides a topological feature that proves that 1L-WS$_2$ A - excitons ($X_A$) couple strongly to SPPs supported by the thin Ag film. The broader B-excitonic ($X_B$) and C- excitonic ($X_C$) resonances couple, on the other hand, only weakly to SPPs. We substantiate our findings by performing additionally traditional angle-dependent reflectivity measurements which yield the dispersion relation of the coupled exciton-SPP system and thus the Rabi splitting and the coupling strength. Analytical transfer matrix method (TMM) as well as complementary numerical Maxwell simulations not only reproduce the experimental findings very well but they also demonstrate that the crossover from the weak to SC is accompanied by a change in the reflection topology.

## 2 Experimental section

1L-WS$_2$ was grown by pulsed thermal deposition (PTD) on sodalime glass as reported previously [35]. The large area (1 cm$^2$) 1L was placed by a polymer-based wet-transfer process on a ~50 nm thick Ag film thermally evaporated on BK7 glass and covered by a 2 nm thick Al$_2$O$_3$ spacer layer fabricated by atomic layer deposition. The Al$_2$O$_3$ film was introduced to protect the Ag film during the wet-transfer process and to avoid direct contact of WS$_2$ with the metal film which is known to deteriorate the optical properties of the 1L. An optical micrograph of the sample is shown in the **Figure S2a of the Supporting Information**. The glass slide was fixed on a cylinder prism by an index matching oil to realize excitation of SPPs in the Ag film in the Kretschmann-Raether configuration [see **Figure 1a**]. The advantage of the semi-cylindrical arrangement is that the light is incident perpendicular to the surface of the cylinder at each goniometer position and therefore no angle correction is needed. The angle of incidence $\theta$ is set with an accuracy of ±0.5° in the experiments. The beam spot diameter of ellipsometer (SENresearch 4.0, Sentech) is ca. 1 mm.

## 3 Results and discussion

### 3.1 TIRE principle

The result of an ellipsometric measurement is expressed by two parameters, the ellipsometric angles $\Delta$ and $\Psi$. They are defined by the ratio $\rho$ of the complex Fresnel reflection coefficients $r_p$ and $r_s$ for p- and s-polarized light, respectively,

$$\rho = \frac{r_p}{r_s} = \tan(\Psi) \cdot e^{i\Delta}.$$

Here, $\tan(\Psi)$ is the amplitude of $\rho$ and provides thus the ratio $|r_p/r_s|$ and $\Delta = \varphi_p - \varphi_s$ is the phase difference between the reflection coefficients of p- and s-polarized light. The generally unintuitive parameters $\tan(\Psi)$ and $\Delta$ can be better understood in the case of TIRE with SPPs with the following considerations. It is well known that SPPs at a metal-dielectric interface can only be excited with p-polarised light [36]. An ellipsometry measurement provides thus the phase and amplitude spectra of the polarization experiencing the resonance relative to the orthogonal polarization which experiences no resonance. The SPP resonance manifests itself as a minimum in the amplitude $|r_p|$ as well as a rapid change in phase $\varphi_p$. Energy and momentum conservation determine thereby the resonance energy of the SPP for a given angle of incidence $\theta$, or a given in-plane wave vector, respectively. In absence of a resonance the reflection coefficient $r_s$ varies only slowly with the photon energy and can therefore be considered as a reference signal to $r_p$. $|r_p|$ is thus proportional to $\tan(\Psi)$ scaled by $|r_s|$ while $\varphi_p$ is equal to $\Delta$ shifted by $\varphi_s$. Representative $\Psi$ and $\Delta$ spectra recorded for an area of the BK7/Ag/Al$_2$O$_3$ structure not covered with WS$_2$ are presented in the **Figure S1 of the Supporting Information**.

### 3.2 TIRE of a planar Ag/Al$_2$O$_3$/1L-WS$_2$ stack to reveal SC

The ellipsometric response of the full BK7/Ag/Al$_2$O$_3$/1L-WS$_2$ structure recorded at an incident angle $\theta = 43.55°$ is depicted in **Figure 1b**. The angle is chosen such, that the SPP propagating at the Ag surface is nearly in resonance with the A-excitonic transition in 1L-WS$_2$. Clearly visible are two minima in the $\Psi$ spectrum at 1.980 eV and 2.046 eV which can be assigned to the coupled $X_A$-SPP resonances as will be substantiated later. Simultaneously, a rapid phase change is observed as the photon energy is tuned through the coupled resonances. Before discussing the $X_A$-SPP coupling, we use the ellipsometric data to retrieve the dielectric function of our PTD-grown 1L-WS$_2$ and the silver film. The fits were performed with the SprectraRay/4 software (Sentech) which is based on the TMM assuming isotropic layers. This simplification might seem questionable on a first glance since the transition dipole moments of the 1L-WS$_2$ $X_A$ and $X_B$ resonances lie in the plane of the monolayer ($x - y$ plane in **Figure 1a**). Numerical simulations of the ellipsometric spectra performed with an isotropic as well as an anisotropic model for the dielectric functions of 1L-WS$_2$ yield very similar results (data not shown). This finding is reasonable since the evanescent electric field of SPPs propagating along the $x -$ direction (Fig. 1a) possesses due to its confinement at the metal-dielectric interface both a transverse ($E_z$) as well as longitudinal component $E_{\mathrm{x}} \approx \pm i E_z \frac{|k_z|}{k_{\mathrm{x}}}$) [37]. Application of Gauß law $\mathrm{div}\vec{E} = 0$ yields $|E_{\mathrm{x}}/E_z| \approx \sqrt{\varepsilon_1/|\varepsilon_2|} \approx 1.3$ with $\varepsilon_1 \approx 9$ and $\varepsilon_2 \approx -16$ being the real parts of the dielectric functions of the Al$_2$O$_3$/1L-WS$_2$ stack and Ag, respectively, in the region of the resonance energy. Therefore, the TMM simulations throughout this paper are performed with isotropic layers. In a first step, we determined the complex dielectric function of the Ag film using the ellipsometry measurement at a sample spot not covered by WS$_2$ (see **Figure S1 of the Supporting Information**). The optical response of the Ag film is described by the Drude-Lorentz oscillator model. Literature data of the dielectric function are used for the BK7 glass and the Al$_2$O$_3$ layer [38, 39]. The obtained parameters of the Ag film are then kept fixed in the fit of the entire BK7/Ag/Al$_2$O$_3$/1L-WS$_2$ stack. 1L-WS$_2$ was modelled by four Tauc-Lorentz oscillators to account for A, B, C, and D excitonic resonances. Details on the fitting procedure and the fitting parameters are given in **section SII of the Supporting Information**. The derived real ($n$) and imaginary part ($\kappa$) of the dielectric function of 1L-WS$_2$ are depicted in **Figure 1c**. The spectra show the A and B excitonic resonances. The oscillator strength of $X_A$ is about 23% smaller than that reported in the literature [40]. Also the obtained thickness of the WS$_2$ layer of 0.72 nm is somewhat smaller than the reported value of 0.8 nm [40]. This might be related to the wet transfer process which deteriorates somewhat the quality of our PTD-grown monolayer. Here, one should bear in mind that the beam spot diameter of the ellipsometer is rather large (ca. 1 mm) and therefore a large area of the 1L-WS$_2$ is interrogated.

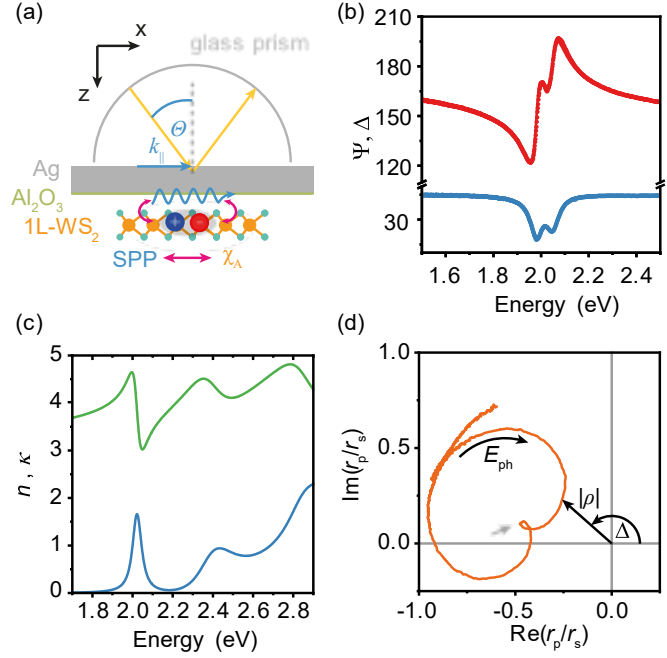

Figure 1: a) Sample layout and TIRE geometry to study the coupling between SPPs of an Ag film and 1L-WS$_2$ excitons. An Ag (50.8 nm)/Al$_2$O$_3$ (2 nm)/1L-WS$_2$ stack is attached to a BK7 glass prism. Ellipsometry measurements are performed at variable angle of incidence $\theta$. b) Ellipsometric parameters $\Psi$ (blue) and $\Delta$ recorded at an angle $\theta = 43.55°$. The symbols represent the experimental data. The solid lines are fits to the data as described in the main text. c) Real ($n$) and imaginary part ($\kappa$) of the dielectric function of 1L-WS$_2$ extracted from the ellipsometry measurement in (a). d) Representation of the phasor $\rho(E_{\mathrm{ph}})$ in the complex plane at an angle $\theta = 43.55°$. The photon energy $E_{\mathrm{ph}}$ covers a range from 1.46 eV to 3.54 eV and runs along the black arrow. The orange arrows in (a) and (c) indicate a photon energy of 2.022 eV.

**Figure 1d** shows an alternative representation of the ellipsometric data of **Figure 1b**. It presents the phasor of $\rho$ in the complex plane with the photon energy $E_{\mathrm{ph}}$ as parameter. A SPP resonance in $r_p$ appears in such a representation as a circle (see **Figure S1b of the Supporting Information**). The simple classical analogy is the complex dissipated power of a harmonic oscillator which yields, when depicted in the complex plane, a circle. The phase offset caused by the "non-resonant" $r_s$ simply leads to a rotation of the circle described by $\rho$ in the complex plane. Most importantly, in the complex plane representation of $\rho(E_{\mathrm{ph}})$ of the present BK7/Ag/Al$_2$O$_3$/1L-WS$_2$ stack one observes not just a simple loop but two intertwined loops which is a fingerprint of the presence of two resonances which are strongly coupled, namely that of the Ag-SPP and the 1L-WS$_2$ $X_A$. The representation of the ellipsometric signal $\rho(E_{\mathrm{ph}})$ in the complex plane provides a means to distinguish between the weak and strong coupling regime of excitons and SPPs as will be corroborated in more detail by TMM simulations below. In a previous work, it has been demonstrated that the phasor representation of $\rho(E_{\mathrm{ph}})$ of the reflection coefficient of an organic microcavity can be used to proof SC of Frenkel excitons with cavity photons [41].

### 3.3 Angle-dependent TIRE and polariton dispersion

Before presenting the simulations we first corroborate our claim of SC between the Ag-SPP and the 1L-WS$_2$ $X_A$ in the simple planar configuration by providing further experimental evidence. We measured $\Delta$ and $\Psi$ spectra of the BK7/Ag/Al$_2$O$_3$/1L-WS$_2$ stack in a range of angles $\theta$ so that the SPR is swept through the 1L-WS$_2$ A-excitonic resonance. Traditionally, angle-dependent measurements of the intensity of the reflected light are performed to prove SC [42]. The angle $\theta$ is related to the in plane wavevector $k_\parallel = \frac{2\pi \cdot n_{\mathrm{BK7}}}{\lambda} \sin\theta$ with $n_{\mathrm{BK7}}$ being the refractive index of the prism and $\lambda$ the wavelength of the incoming photons. Variation of $\theta$ yields thus the dispersion of the coupled $X_A$-SPP system. **Figure 2a** shows a density plot of the $|r_\mathrm{p}/r_\mathrm{s}|$ spectra vs. the angle $\theta$. Individual $\Delta$ and $\Psi$ spectra for selected angles $\theta$ are reported in the **Figure S2 of the Supporting Information**. The yellow regions in **Figure 2a** correspond to minima in $|r_p/r_s|$ caused by the photoexcitation of the coupled $X_A$-SPP resonances. They represent the high- and low-energy polaritonic branches. At small energies and angles, a linear relationship is observed. It corresponds to the plasmonic part of the excitation. The excitonic part represents a horizontal line in such a plot. Toward larger angles, as the SPP energy approaches the 1L-WS$_2$ $X_A$ resonance energy, repulsion and anticrossing of the two branches is observed which is indicative of SC between the two resonances. TMM simulations of $|r_p/r_s|$ of the Ag/Al$_2$O$_3$/1L-WS$_2$ stack embedded between BK7 glass and air reproduce the experimental data very well [Fig. 2(b)]. The simulations, performed with the dielectric functions of the Ag film and the 1L-WS$_2$ retrieved from the ellipsometry measurements at one fixed angle, reproduce the experimental angle dispersion of the two polariton branches. To further support our conclusion of SC between 1L-WS$_2$ $X_A$ and Ag-SPP we performed also numerical calculations which reproduce the TMM simulations and show furthermore that the splitting is not only observed in the $\Psi$ resp. $|r_p/r_s|$ spectra but also in the absorption of the coupled system as well as in the constituents (see **Figure S5 of the Supporting Information**). Therefore, we can safely conclude that the observed dips in reflectivity do not stem from enhanced absorption but are a signature of SC.

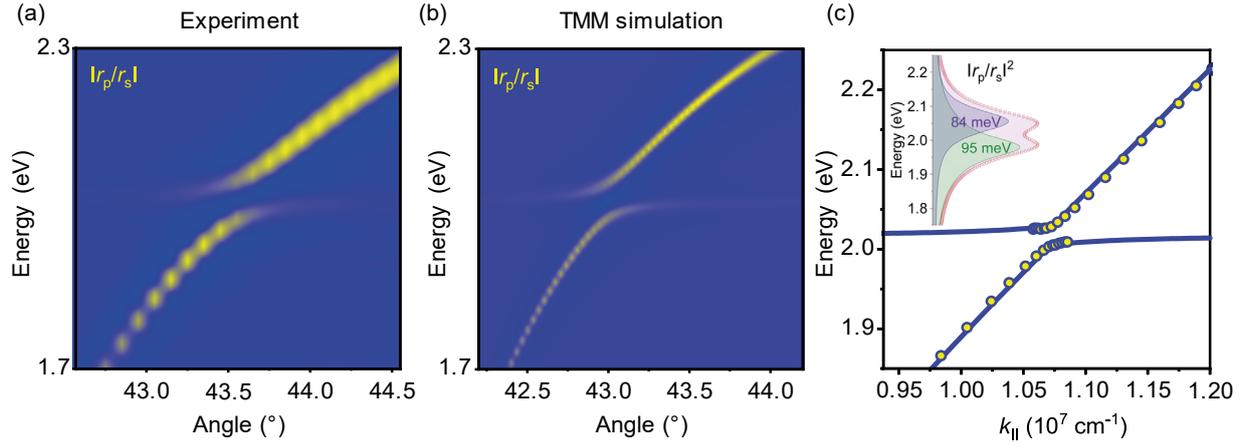

Figure 2: Density plots of the $|r_p/r_s|$ spectra of the Ag/Al$_2$O$_3$/1L-WS$_2$ stack embedded between BK7 glass and air vs. the angle $\theta$ obtained from ellipsometry measurements (a) and TMM simulations (b). For better contrast, the second derivative $\frac{\partial^2}{\partial E^2}|r_p/r_s|$ of the spectra is plotted. The TMM simulations are performed with the dielectric functions of Ag and 1L-WS$_2$ derived from the fit of ellipsometry measurements depicted in Fig. 1(a). c) Dispersion relation $E$ vs. $k_\parallel$ of the coupled 1L-WS$_2$ $X_A$ – Ag SPP resonances. The symbols represent the energetic positions of the minima in the $|r_p/r_s|$ spectra. The solid line is a fit to the data using the coupled oscillator model as described in the main text. The inset shows that the $|r_p/r_s|^2$ spectrum of the coupled $X_A$ – SPP resonances is composed of two peaks with very similar FWHM. The fit of the spectrum was performed with two pseudo Voigt profiles.

In order to derive the magnitude of the Rabi splitting $\hbar\Omega_R$, the photon energies corresponding to the minima in the $|r_p/r_s|$ spectra as a function of the in-plane wavevector $k_\parallel$ are plotted in Fig. 2(c).

The complex eigenenergies of the lower ($E_-$) and upper ($E_+$) polariton branch are obtained from the coupled oscillator model [42] which yields

$$E_\pm = \frac{(E_{X_A} + E_{SPP})}{2} - i\frac{(\gamma_{X_A} + \gamma_{SPP})}{2} \pm \sqrt{(\hbar\Omega_R)^2 + (E_{X_A} - E_{SPP})^2 - \frac{(\gamma_{X_A} - \gamma_{SPP})^2}{4} - i(E_{X_A} - E_{SPP})(\gamma_{X_A} - \gamma_{SPP})}$$

Here, $E_{X_A}, E_{SPP}$ and $\gamma_{X_A}, \gamma_{SPP}$ are the energies and broadening parameters of the uncoupled $X_A$ and SPP resonances, respectively. At resonance, i.e. $E_{X_A} = E_{SPP}$, the complex eigenenergies $E_\pm$ exhibit a splitting in their real parts whenever $\hbar\Omega_R > |\gamma_{X_A} - \gamma_{SPP}|/2$. The imaginary parts remain unsplit which means that both polariton branches decay with a common rate at resonance. This is the so-called SC regime. On the other hand, whenever $\hbar\Omega_R < |\gamma_{X_A} - \gamma_{SPP}|/2$ the real parts of $E_\pm$ coalesce and simultaneously a splitting in the imaginary parts arises which implies a modification of the decay rates. This is the weak coupling regime. This criterium is often used to differentiate between weak and strong coupling regime [43]. It should be noted that there are also other criteria proposed in literature [44]. In the present experiments, we clearly observe a splitting in the real parts of $E_\pm$. The experimental FWHM of the uncoupled $X_A$ and SPP resonances yield broadening parameters $\gamma_{X_A} = 51$ meV and $\gamma_{SPP} = 146$ meV (see **section IV of the Supporting Information**). Notably, when the $X_A$ and SPP transitions are tuned into resonance, the two polariton branches assume a similar linewidth as expected in the SC regime (inset of **Figure 2c**). The fit of the experimental dispersion plot with the coupled oscillator model yields a Rabi splitting of 52 meV (**Figure 2c**). This is larger than the half width at half maximum (HWHM) of the coupled resonances, i.e. $\frac{\gamma_{X_A} + \gamma_{SPP}}{4}$, which is a more intuitive criterium for strong coupling. The coupling strength is $g = \frac{\hbar\Omega_R}{2} = 26$ meV.

At a first glance, the coupling strength seems not very large. However, one has to put into perspective the thinness of the layer in which the WS$_2$ excitons reside (< 1nm) and the effective mode length $L_z$ of the SPP. The latter is a measure of the extension of the electric field $E$ of the SPP which falls off exponentially perpendicular to the interface. A common definition of $L_z$ is $L_z w_{E,max} = \int_{-\infty}^{\infty} w_E dz$ with $w_E$ being the electric field energy density [45]. For a pristine Ag surface in air $L_z \approx$ 225 nm at 2 eV is obtained. The coupling strength is given by $g = \sqrt{N}\mu E$, with the transition dipole moment $\mu$ of each one of $N$ WS$_2$ $X_A$ excitons interacting with the electric field $E$ of the SPP. To obtain an estimate of $g$, the vacuum SPP field at the resonance frequency $\omega$ can be defined by $E = \sqrt{\hbar\omega/2\varepsilon\varepsilon_0 V}$ in analogy to the field in a photonic cavity with the mode volume $V$ [44]. In the present planar configuration, a reasonable definition of the mode volume is $V = L_x L_y L_z$ where $L_z$ is the just introduced effective mode length and the product $L_x L_y$ represents the in-plane coherence area of the plasmonic mode. Introducing the 2D exciton density $n_{2D}$ in 1L-WS$_2$ yields $\sqrt{N/V} = \sqrt{n_{2D}/L_z}$ and, consequently, $g = \sqrt{n_{2D}\mu^2 \hbar\omega/2\varepsilon\varepsilon_0 L_z}$. Considering that the imaginary part $\kappa(\omega)$ of the dielectric function of 1L-WS$_2$ multiplied by the thickness $d$ of the monolayer is proportional to $n_{2D}\mu^2$ yields the simple relation $g = \sqrt{(3/2)\hbar^2\omega\gamma_{X_A}\kappa d/L_z}$ (see **section V of the Supporting Information**). With our experimental values for $\gamma_{X_A}$ and $\kappa$ at 2 eV a coupling strength $g \approx 28.5$ meV is calculated which is close to the value derived from the angle-dependent ellipsometry measurements. Using values for $\gamma_{X_A}$ and $\kappa$ reported in the literature for a CVD-grown of 1L-WS$_2$ [40] yields a larger value $g \approx 40$ meV. As already pointed out, the wet transfer leads to a partial damage of the monolayer resulting in apparent smaller values for $\kappa$ and $d$ as reported in [40]. Due to the smallness of the factor $\sqrt{d/L_z}$, it is surprising that SC of Ag-SPPs with excitons in the atomically thin monolayer can be reached at all in a simple planar geometry. However, the large transition dipole moment of about $\mu = 56$ Debye [46] and the narrow linewidth of the 1L-WS$_2$ of $X_A$ transition make it possible. In contrast, no clear signatures for the SC regime are observed in the angle-dependent $\Psi$ and $\Delta$ spectra in the spectral regions of the 1L-WS$_2$ $X_B$ and $X_C$ transitions (see **Figure S3 the Supporting Information**). Although these resonances possess also large oscillator strengths their broad spectral widths prohibit SC with the Ag-SPP.

### 3.4 Phasor representation of $\rho$ and criterium for SC

Having confirmed that indeed SC between the 1L-WS$_2$ $X_A$ and the Ag-SPP is achieved, we finally corroborate our claim that the occurrence of a secondary intertwined loop in the representation of the phasor $\rho = r_p/r_s$ in the complex plane is a fingerprint of the SC regime. To that end we performed TMM simulations modelling the dielectric response of the 1L-WS$_2$ $X_A$ by Lorentzian oscillators with resonance frequency $\hbar\omega_0 = 2$ eV and broadening parameter $\gamma_L = 50$ meV residing in a thin (0.8 nm) layer. The Ag film is described by the Drude model with a plasma frequency $\hbar\omega_P = 9.2$ eV and a damping term of 26 meV [47]. $|r_p/r_s|$, $\Delta$ and $\rho$ are calculated as a function of the transition strength $\beta$ of the oscillators (Fig. 3). The angle of incidence is fixed to a value so that the Ag-SPP is in resonance with the oscillators. Increasing $\beta$ from 0 to 0.25 drives the system from the weak to the strong coupling regime in the present configuration. At small $\beta \leq 0.15$, the two polariton branches just start to split (see **Figure 3a**), a point of inflection appears in the phase signal $\Delta$ (see **Figure 3b**) and an indentation appears in the curve the phasor describes in the complex plane at $E_{ph} = \hbar\omega_0$ (see **Figure 3c**). As pointed out above, the phasor $\rho$ of a sole SPR circumscribes a circle in the complex plane (see **Figure 3c**). The presence of an absorbing medium on top of the Ag film modifies the internal damping and thus the depth of the reflectivity minimum [48] causing the observed deformation. This is characteristic of the weak coupling regime. If $\beta$ is increased ($\beta > 0.15$), a cross over to the strong coupling regime occurs. While the splitting in the $|r_p/r_s|$ spectra just increases (see **Figure 3a**), the changes of the phase signal $\Delta$ (see **Figure 3b**) and the phasor $\rho$ (see **Figure 3a**) are more profound: The point of inflection in the $\Delta$ signal develops into a S-shape, i.e. the steep phase jump of a single resonance evolves into two subsequent phase jumps which indicates the excitation of two coupled resonances. Even more striking is the change in the topology of $\rho$ (see **Figure 3c**). A double point appears as the single loop transforms into two intertwined loops corroborating the existence of two coupled resonances. In the representation of $\rho$ in the complex plane the cross over from the weak to the strong coupling regime can thus be pinpointed more firmly than by measuring the splitting in the reflection spectrum. In the present configuration, SC sets in at $\beta \approx 0.2$. The corresponding splitting in the $|r_p/r_s|$ spectrum is about 40 meV. The FWHM of the Lorentz oscillators is $\gamma_L = 50$ meV. TMM simulations of $|r_p/r_s|^2$ of the pristine Ag (50 nm)/Al$_2$O$_3$(2 nm) stack yield a FWHM $\gamma_{SPP} = 107$ meV of the uncoupled SPR of the Ag/Al$_2$O$_3$ stack (data not shown). As pointed out above, in the SC regime the split resonances decay with a common rate, i.e. $\gamma_{SPP-L} = (\gamma_{SPP} + \gamma_L)/2$. The simulations imply, that a double point in the phasor $\rho$ and thus SC sets in when the splitting of the resonances exceeds the half width at half maximum $\gamma_{SPP-L}/2$, i.e. $\frac{(\gamma_{SPP}+\gamma_L)}{4} \approx 39$ meV in agreement with our experimental observation.

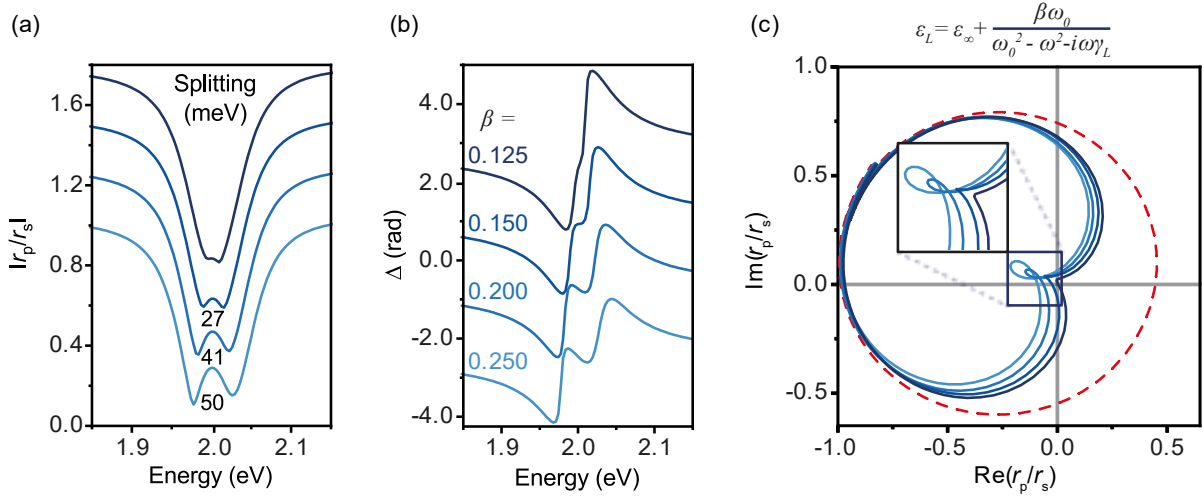

Figure 3: TMM simulations of an Ag (50 nm)/Al$_2$O$_3$(2 nm)/Lorentz oscillator (0.8 nm) stack embedded between BK7 glass and air. As in the experiment, the light is incident from the BK7 glass. The ellipsometric parameters $\tan\Psi = |r_p/r_s|$ (a) and $\Delta$ (b) as well as the parametric plot of $\rho(E_{ph})$ in the complex plane (c) are calculated as functions of the strengths $\beta$ of the oscillators given in (b) (different shades of blue). The definition of $\beta$ is given in (c). For comparison, the red dashed curve in (c) depicts $\rho(E_{ph})$ of the uncoupled SPP obtained by setting $\beta = 0$.

**4 Conclusion and outlook**

We have shown that SC of SPPs of a thin Ag film with WS$_2$ $X_A$ excitons can be achieved even for WS$_2$ in the monolayer regime and in an utmost simple planar geometry. This is due to the large transition dipole moment and narrow linewidth of the 1L- WS$_2$ $X_A$ transition. The observation of SC is possible due to the large area WS$_2$ monolayer deposition onto a flat silver film which is a prerequisite to study the system by means of TIRE. We have shown that this method provides a distinct fingerprint of the strong coupling regime. The planar hybrid geometry can be considered as a building block for future active plasmonic devices. Such applications require coupling to SPPs propagating in plasmonic waveguides [11, 14] which simplifies the sample preparation significantly since only a few tens of μm-size well-defined monolayer flakes are needed. Well-established hBN encapsulation obtained by dry transfer techniques is known to improve the optical properties significantly and excitonic linewidths down to 24 meV have been reported [49, 50]. Higher coupling strengths are thus within reach with more optimized monolayer quality. Dielectric engineering in order to confine the electric field more strongly at the metal-dielectric interface could be employed to further increase the coupling strength $g \sim \sqrt{1/L_z}$ and make the SC more robust. The well-known susceptibility of the 1L-TMDC optical properties towards external stimuli like electric fields and light, as well as the dielectric/chemical environment renders active control over the light matter interaction and even tuning the system between the weak and strong coupling regime possible. This would open up the opportunity to study the physics of exceptional points [51] in 2D-exciton-SPP systems.

**Supporting Information**

Spectral ellipsometry of an Ag/Al$_2$O$_3$ reference stack; extraction of the dielectric function of Ag and of 1L-WS$_2$; optical microscope image of sample; angle-dependent TIRE spectra covering the spectral range of the 1L-WS$_2$ $X_A$, $X_B$ and $X_C$ transitions; linewidth of Ag SPP; derivation of the equation for the coupling strength; numerical simulations.

**ACKNOWLEDGMENT**

The authors gratefully acknowledge financial support by the Deutsche Forschungsgemeinschaft through CRC 951 (Project number 182087777). Sergey Sadofev acknowledges financial support by the Deutsche Forschungsgemeinschaft (DFG) through the grant SA 3039/1-1.